\documentclass[conference]{IEEEims} 
\input epsf
\usepackage{graphicx}
\begin{document}

\title{\LARGE Ballistic  transport and electrostatics 
in metallic carbon nanotubes}

\author{\authorblockN{A. Svizhenko\authorrefmark{1}, 
M. P. Anantram\authorrefmark{2} and T. R. Govindan\authorrefmark{3}}
\authorblockA{ Center for Nanotechnology and NASA Advanced Supercomputing Division,\\
NASA Ames Research Center,\\
Mail Stop: \authorrefmark{1}229-1 \authorrefmark{2}229-1
              \authorrefmark{3}258-5}
Moffett Field, CA 94035-1000 \\
Email: \authorrefmark{1}svizhenk@nas.nasa.gov 
\authorrefmark{2}anant@mail.arc.nasa.gov  
\authorrefmark{3}Tr.R.Govindan@nasa.gov}


\maketitle

\begin{abstract}

We calculate the current and electrostatic potential drop in metallic
carbon nanotube wires self-consistently, by solving the Green's 
function and electrostatics equations in the ballistic case.
About one tenth
of the applied voltage drops across the bulk of a nanowire, independent
of the lengths considered here.
The remaining nine tenths of the bias drops near the contacts, thereby
creating
a non linear potential drop. 
The scaling of the 
electric field at the center of the nanotube with length (L) is faster
than $1/L$ (roughly $1/L^{1.25-1.75}$). 
At room
temperature, the low bias conductance of large diameter nanotubes is
larger than $4e^2/h$ due to occupation of non crossing subbands.
The physics of conductance evolution with bias due to the 
 transmission Zener tunneling in non crossing subbands is discussed.
\end{abstract}

\IEEEoverridecommandlockouts

%
\IEEEpeerreviewmaketitle

\section{Introduction}
Carbon nanotubes have the potential
to make good interconnects because current densities approaching $10^9
\; A/cm^2$~\cite{wei-apl-01} and compatibility with silicon 
technology~\cite{kreupl-microeleceng-02,ngo-eeetn-04} have
been demonstrated. In the use of multiwall nanotubes for such applications, 
contact with as many layers of the tube as possible~\cite{pablo-apl-99} is desirable.  Larger 
diameter nanotubes are preferable because even the semiconducting shells
can carry current due to small bandgaps. For example, a 
semiconducting shell with a radius of 18.8 nm  has a half bandgap of
only 62.5 meV (kT=26 meV at room temperature).
Experimentally measured conductance of metallic carbon nanotubes
is close to the theoretical maximum of $4e^2/h$ at low
biases~\cite{frank-science-98,poncharal-epjd-99,nygard-applphysa-99,
kong-prl-01}.

Experiments on small diameter carbon nanotubes show that the 
differential conductance decreases with applied bias, for voltages
larger than 100 - 200 meV~\cite{yao-prl-00,collins-prl-01,pablo-prl-02}.
Ref. \cite{yao-prl-00} found that the conductance 
decrease with increase in bias was caused by reflection of incident electrons 
at crossing subbands due to scattering with zone boundary 
phonons. Measured values of the mean free path ~\cite{park-nl-04}
are in the range of 0.1 - 1 $\mu$m at low bias and around 10 nm at high bias for
1.8 nm diameter nanotubes (corresponds to a (24,0) nanotube).
Simple estimates of scattering rates  ~\cite{park-nl-04,golgsman-prb-03}
show inverse dependence with nanotube diameter, which suggests that 
phonon scattering is less important in large diameter nanotubes.

The non crossing subbands of small diameter nanotubes do not 
carry current due to their large band gap. In contrast, large diameter 
nanotubes experimentally show an increase in conductance with applied
bias~\cite{frank-science-98,poncharal-epjd-99,poncharal-jpcb-02,liang-apl-04}. 
Reference \cite{Anantram-prb-00} suggested that as the diameter 
increases electrons may tunnel into non crossing subbands, thus 
causing an increase in differential conductance with applied bias. The
main drawbacks of the calculation in reference \cite{Anantram-prb-00}
was that the results depended on the assumed form of potential drop in
metallic nanotubes (charge self-consistency and phonon scattering were
neglected).

Apart from measuring current-voltage characteristics, the potential
drop can be measured by EFM. Refs.  \cite{bachtold-prl-00} found that
the potential drop in a single wall nanotube was small compared to a 
multiwall nanotube.

In this work, we computationally model ballistic electron flow in metallic
nanowires within a tight-binding approximation by  including {\it charge 
self consistency}, which is an important factor in 
determining the current. 
 The potential distribution and current-voltage characteristics are studied
as a function of both nanotube diameter and length.
In 3D (bulk) crystal the electric field, potential and net charge at the edges of the tube 
would decay exponentially, resulting in zero net charge and 
zero electric field inside the conductor.
In a 1D wire, even perfectly ballistic,  
the screening is poorer, due to the lower electron density of states.
Reference \cite{leonard} found that near {\it p-n} junctions in 
semiconducting nanotubes, this 
leads to a very long range ($\sim 1/ln(y)$) 
tail in the charge distribution 
and, consequently,  
a non zero electric field very far from the junction. 
It is however an open question whether metallic 
nanotubes also have poor screening properties. 
Here we use an electrostatics model similar to that in \cite{leonard}, 
but significantly improve on 
transport, using Non Equilibrium Green's Function formalism.
The ballistic approximation works well at 
low bias and/or in large diameter nanotubes. 
The detailed study of phonon scattering effects will be given elsewhere.

The nature of the metallic contacts is also
important. From an experimental view point the contact between a metal and 
a nanotube can either be an end-contact or side-contact. 
The end-contact
 corresponds to only the nanotube tip making the contact to the 
  metal lead. In experiments, the end-contacts 
   usually include strong chemical 
modification of the nanotube at the metal-nanotube interface.
  Also, reference \cite{frank-science-98} found that end-contacts without 
  sufficient chemical modification of the nanotoube-metal interface
   have a large contact resistance.
    In such configuration, 
   electrons are injected directly from the metal to the nanotube 
   region between the contacts. 
     Due to the uncertainty of the contact bandstructure,  
 modeling end-contacts even remotely
  correct is extremely difficult.
   
   The side-contacts correspond to 
   coupling between metal and nanotube atoms over many unit cells
    of the nanotube, and can be thought of as a nanotube buried 
    inside a metal. Most experimental configurations correspond to
     side-contacts \cite{frank-science-98,tans}. 
     An important feature of the side contacts is that the  coupling between
      atoms in the nanotube is much stronger than coupling between 
      nanotube and metal atoms which means that the bandstructure 
      of the contacts can be assumed to be
       the same as that of the nanotube between the contacts.     
     In the side-contact geometry, 
      electrons are predominantly injected from the metal into the
       nanotube buried in the metal and then transmitted to the 
       nanotube region between the contacts.              
        In fact, as proof of such a process, scaling 
       of conductance with contact area has been observed in the 
       side-contacted geometry by references \cite{frank-science-98,tans}. 
       Modeling showed that the conductance in metallic zigzag 
       nanotubes can be close to the theoretical maximum for 
       conductance with sufficient overlap area in the 
       contact \cite{Anantram-apl-01}.

In the work presented here, we assume  "perfect" contacts to 
represent an ideal side-contacted nanotube.  The details of the contact model
will be given below.

\section{Formalism}

\begin{figure}
\epsfxsize=3.25in\epsfbox{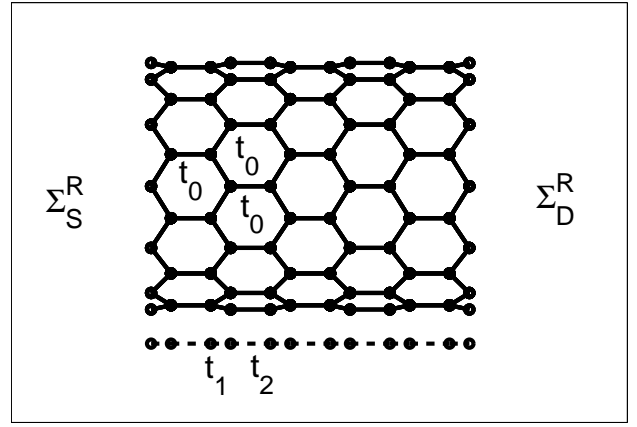}
\caption{Zigzag carbon nanotube and the corresponding 1D chain. 
The hopping parameter between nearest neighbors in the nanotube is $t_o$. 
The 1-D 
chain has two sites per unit cell
with on-site potential $U_o$ and 
hopping parameters $t_1=2t_o
\cos({\tilde qa \over 2})$ and $t_2=t_o$ }
\label{fig:tube}
\end{figure}

In this paper we consider only zigzag carbon nanotubes. 
The analysis for armchair nanotubes is similar.
The general form of the Hamiltonian for electrons in a carbon nanotube
 can be written as:
\begin{eqnarray}
H & = & \sum_{i,x}U^{i}_{x}c^{\dagger}_{i,x}c_{i,x}+
\sum_{i,j,x,x'}
t^{i,j}_{x,x'}c^{\dagger}_{i,x}c_{j,x'}
\end{eqnarray}
The sum is taken over all rings $i,j$ in transport direction and all atoms located at $x,x'$ 
in each ring. 
We make the following common approximations:
i)  only nearest neighbors are included; 
each atom in an $sp^2$-coordinated carbon nanotube
has three nearest neighbors, located $a_{cc}=$ 1.42 $\AA$ away;
ii) the bandstructure consists of only $\pi$-orbital, with the 
hopping parameter
$t_o=V_{pp\pi}=-2.77$ eV and the on-site potential $U_o=\epsilon_p=0$. 
Such a tight-binding model 
is adequate to model transport properties in undeformed nanotubes.
Within these approximations, only the following parameters are non zero:
\begin{eqnarray}
U^{i}_{x}& = & U_o , \forall i \nonumber \\
t^{i,i-1}_{x,x'}& = &t^{i-1,i}_{x,x'}=t_o\delta_{x\pm a/2,x'}, \forall i=2k 
\label{eq:hop_par}  \\
t^{i,i+1}_{x,x'}& = &t^{i+1,i}_{x,x'}=t_o\delta_{x,x'}, \forall i=2k ~,
\nonumber
\end{eqnarray}
where $a=a_{cc}\sqrt{3}$.
 In $(N,0)$ zigzag nanotubes, the 
wave vector in the circumferential direction is quantized as 
$\tilde q=2\pi q/Na$, $q=1,2,...N$, 
creating eigenmodes in the energy spectrum. 
By doing a Fourier expansion of $c^{\dagger}_{i,x}$ and $c_{i,x}$ 
in $\tilde q$-space  and using  (\ref{eq:hop_par}) we obtain
a decoupled electron Hamiltonian in the eigenmode space:
\begin{eqnarray}
H & = & \sum_{q} H^{q}\\
H^{q}& = & \sum_{i}
(U^{i}_{\tilde q}
c^{\dagger}_{i,\tilde q}c_{i,\tilde q}
+t^{i,i\pm1}_{\tilde q}
c^{\dagger}_{i,\tilde q}c_{i\pm1,\tilde q})
\end{eqnarray}
where
\begin{eqnarray}
U^i_{\tilde q} &=& U_o, \forall i \nonumber\\
t^{i,i-1}_{\tilde q} & = & t^{i-1,i}_{\tilde q}=2t_o
\cos({\tilde qa \over 2}) \equiv t_1, \forall i=2k  \\
t^{i,i+1}_{\tilde q}& = & t^{i+1,i}_{\tilde q}=
t_o \equiv t_2,  \forall i=2k  \nonumber
\end{eqnarray}
The 1-D tight-binding Hamiltonian $H^q$ describes  a  
chain with two sites per unit cell
with on-site potential $U_o$ and 
hopping parameters $t_1$ and $t_2$ (Fig. \ref{fig:tube}).
For numerical solution, the spatial grid corresponds to the rings of the nanotube, 
separated by $a/2$ with a unit cell 
length of $3a/2=$ 2.13 $\AA$, which is half the unit cell length of a 
zigzag nanotube. 
 
The subband dispersion relations are given by
\begin{eqnarray}
E_q(k)=\pm {\bigl| 1+4cos({3ak\over2})cos({q\pi\over N})+
4cos^2({q\pi\over N})\bigr|}^{1/2}
\end{eqnarray}
Therefore, when $N=3k$,
 there are two subbands with zero bandgap 
 the tube is metallic. In the rest of the paper we  
 distinguish between metallic 
or {\it crossing} subbands  ($q=N/3$ and $2N/3$) and semiconducting or 
{\it non crossing} subbands.

 The contacts 
are assumed to be reflectionless reservoirs
 maintained at equilibrium, i.e. they have  well defined chemical 
potentials, equal to that of the metal leads: $V_S$ in the source and $V_D$  in 
the drain.
Further, the nanotube and metal
 are assumed to have the same workfunction.
The contact self-energies $\Sigma_{c,q}^{R,<,>}=
\Sigma_{S,q}^{R,<,>}+\Sigma_{D,q}^{R,<,>}$ due 
to the source contact are found
\cite{JAP} using
the surface Green's function $g^{S}_q$ of a semi-infinite nanotube, 
which is the solution 
of the following system of equations:
\begin{eqnarray}
(a_1-t_{2}^2g^{S}_{2,q})g^{S}_{1,q}&=&1 \nonumber\\
(a_2-t_{1}^2g^{S}_{1,q})g^{S}_{2,q}&=&1 ~, \label{eq:surf_gr}
\end{eqnarray}
where the indices $1$ and $2$ stand for the two sites of the unit cell and
$a_{1,2}=E-U_o-V_{S}$. The Green's function for the drain
are solved for in a similar way, by making the following substitutions:
$t_{1,2}^{S} \rightarrow t_{2,1}^{D}$, 
$g_{1,2,q}^{S}\rightarrow g_{1,2,q}^{D}$.

For each subband $q$ we solve a system of transport equations
 \cite{JAP}:
\begin{eqnarray}
(E-H^q-V-\Sigma_{c,q}^{R}(E))G^{R,q}=I \label{eq:tran1}\\
(E-H^q-V-\Sigma_{c,q}^{R}(E))G^{<,>,q}=
\Sigma_{c,q}^{<,>}(E)G^{A,q} 
\label{eq:tran2}
 \end{eqnarray}
Electron charge  and current density $n_i$ and $J_i$  at each node $i$ 
are found from the following equations:
\begin{eqnarray}
n_i=-{2i}\sum_{q} 
\int_{-3t_o-eV_D}^{-eV_S+10kT}G^{<,q}_{i,i}(E){dE \over 2\pi} \label{eq:dens} \\ 
J_i={4e \over \hbar }\sum_{q} \int_{-eV_D-10kT}^{-eV_S+10kT}
G^{<,q}_{i,i+1}(E)t^q_{i+1,i}{dE \over 2\pi}, 
\end{eqnarray} 
where we used the fact that the band 
bottom at equilibrium is at $-3t_o$.  
To integrate the electron charge  over energy,
we employed a conventional approach 
developed in DFT (\cite{Williams,Guo,Stokbro}) and described below.
The technique  
drastically reduces the computational requirements as the whole 
band can be integrated with 200 energy grid points with error 
of $10^{-6}$ of charge.
The integral (\ref{eq:dens}) is divided into equilibrium $[-3t_o-eV_D; -eV_D-10kT]$ and 
non-equilibrium parts $[-eV_D-10kT, -eV_S+10kT]$.  In the equilibrium part 
of the energy spectrum, the current is zero and one can apply the 
fluctuation-dissipation theorem $G^<(E)=-2if(E)\Im m[G^R(E)]$,  where 
Fermi factor $f(E)=1$ at these energies. 
$G^R$ is analytical in the upper plain of 
complex energies which allows 
one to transform the integration of $G^<$  over the real axis in the 
equilibrium part of energy spectrum to the integration of $G^R$ over a complex contour $C$ 
starting at $-3t_o-eV_D$ and ending at $-eV_D-10kT$. 
The non equilibrium part is integrated over the real axis from 
 $-eV_D-10kT$ to $-eV_S+10kT$:
\begin{eqnarray}
n_i&=& -{2 \over \pi}\sum_{q} \oint_C\Im m[G^{R,q}_{i,i}(E)dE] \nonumber \\
&&-{i \over \pi}\sum_{q} \int_{-eV_D-10kT}^{-eV_S+10kT}G^{<,q}_{i,i}(E)dE 
\label{eq:densint}
\end{eqnarray}
 
In this work, corresponding to ballistic transport, we neglect electron-phonon scattering. 
The electron mean free path (mfp) is proportional to the matrix element squared and thus 
inversely proportional to the chirality index.
 Ref. \cite{park-nl-04} reported 
$\lambda_{o}^{el} \approx$ 1.6 $\mu m$ and 
$\lambda_{o}^{in}\approx$ 10 nm for a tube 
with a diameter of 1.8 nm, corresponding to a (24,0) nanotube.
This  allows us to assume that large diameter nanotubes of 
moderate lengths, e.g. $\sim\,$100 nm (240,0) nanotube, will remain nearly ballistic even at high bias.  
The ballistic transport is also an adequate approximation at low bias for all diameter nanotubes due to the absence
of strong inelastic phonon emission.

We model the electrostatics of the nanotube as a system of 
point charges  between the two contacts located at $y=y_S=0$ and 
$y=y_D=L$.
The "perfect contacts" are modeled as parallel semi-infinite 
three dimensional
 metal leads that are maintained at fixed source and drain potentials: 
 $V_S$ for $y<y_S$ and $V_D$ for $y>y_D$. 
 So, while the self-energies due to contacts are identical to that 
 of a semi-infinite nanotube, the role of electrostatics is included
  by image charges corresponding to a perfect metal. 
 The electrostatic potential consists of  a linear drop due to a 
uniform electric field
created by the leads and the potential due to the charges on the 
tube and their images
\begin{eqnarray}
V=-eV_S-e(V_D-V_S)(y-y_S)/(y_D-y_S)\nonumber \\
+\sum_{j}G(i,j)(n_j-N) \label{eq:prof}
\end{eqnarray}
with the Green's function  
{\small \begin{eqnarray}
G(i,j)&=&{e\over 4N\pi\epsilon_o}\sum_{n=-\infty}^{+\infty} \sum_{l}
\biggl[ {1\over\sqrt{(y_i-y_j+2nL)^2+\rho_{k,l}^2}} \nonumber\\
& & \;\;\;\;\;\;\;\;\;\;\;\;\;\;\;\;
-{1\over\sqrt{(y_i+y_j+2nL)^2+\rho_{k,l}^2}}\biggr]~.  \label{eq:green}
\end{eqnarray}}
Here,  $\rho_{k,l}$ is the radial projection of the vector between atom 
$k$ at ring $i$ and atom $l$ at ring $j$. The summation is 
performed over all atoms $l$ at ring $j$ for an arbitrary value of $k$.
Maintaining 
the nanotube atoms buried in the metal at a fixed potential is close 
to reality because of the screening properties of 3D metals. Within 
a few atomic layers from the metal surface, the potential should have
 approached the bulk values. While the variation in potential in these
  few atomic layers of the 3D metal is not captured in our model, our
   conclusions on the nanotube electrostatics should not be significantly 
   affected. 

The calculations for one bias point involve 
iterations of (\ref{eq:surf_gr}-\ref{eq:prof}) 
until convergence 
of the
potential and charge distributions is achieved.
The solution of (\ref{eq:tran1}-\ref{eq:tran2}) employs 
the recursive algorithm \cite{JAP}, which scales linearly  
with the number of nodes. 
The contour integral in (\ref{eq:densint})
is taken using Gaussian quadratures.

\section{Results}

\subsection{Electrostatics}

\begin{figure}
\epsfxsize=3.25in\epsfbox{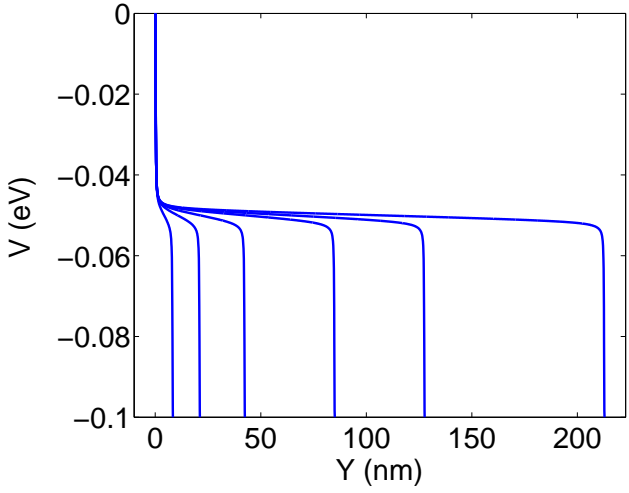}
\caption{Potential distribution versus position for (12,0) nanotubes
of various lengths. The diameter of the (12,0) nanotube
is 0.94 nm and the applied bias is 100 mV. Note the constant potential drop at the edges. }
\label{fig:pot1}
\end{figure}

At low biases (100 mV), electron-phonon scattering does not play a significant
role in determining the potential distribution for wires of moderate length
(less than few hundreds of nanometer). The potential distribution for 
(12,0) nanotubes of 
lengths varying from  21.3 to 213 nm are shown in Fig. \ref{fig:pot1}.
The edges of the nanotube near the contact rapidly screen the applied
bias / electric field.  
The screening is due to the free electron density provided by the crossing subbands.
The potential drop is divided unequally
between different parts of the nanotube,
 with 90\% of the applied bias dropping within 
1 nm from the edges for both lengths considered here.  
Note, that in the ballistic regime, 
resistance of the nanotube does not depend on tube length, 
which results in the drop at the edges being constant for all tube lengths. 

\begin{figure}
\epsfxsize=3.25in\epsfbox{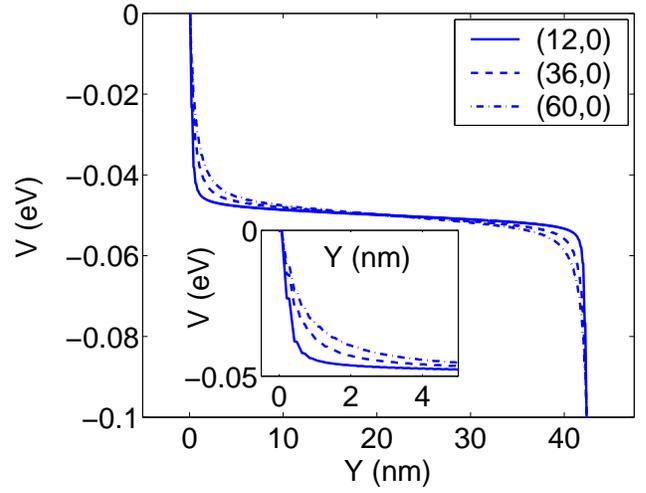}
\caption{Potential versus position for (12,0), (36,0) and (60,0) nanotubes, 
with diameters 0.94, 2.82 and 4.7 nm. The screening for the large
diameter nanotubes is poorer. The inset magnifies the
potential close to the edge of the nanotube. The applied bias is
100 mV.}
\label{fig:pot2}
\end{figure}
\begin{figure}
\epsfxsize=3.25in\epsfbox{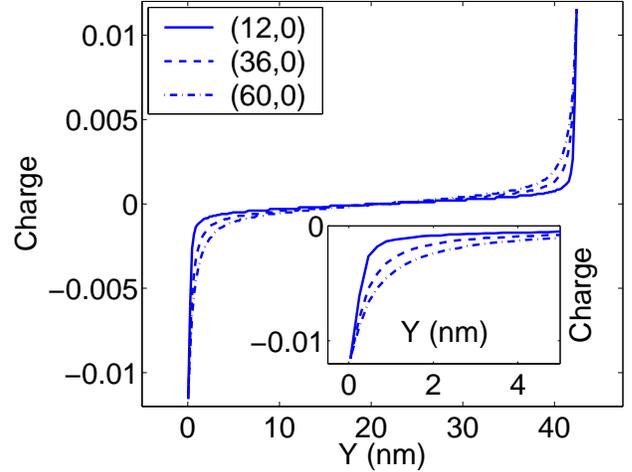}
\caption{Charge density versus position for (12,0), (36,0) and (60,0) nanotubes.
The inset magnifies the
charge distribution close to the edge of the nanotube. The applied bias is
100 mV.}
\label{fig:dens2}
\end{figure} 
\begin{figure}
\epsfxsize=3.25in\epsfbox{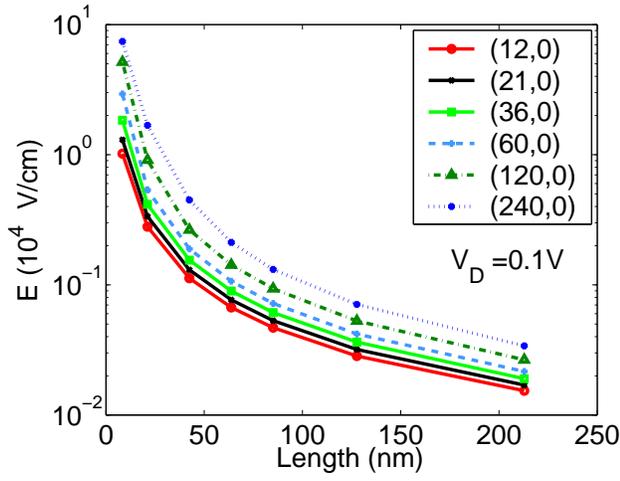}
\caption{This plot shows the electric field at the mid point versus 
nanotube length for various nanotube diameters at an applied bias of 100 mV.
The electric field decreases approximately as 
$1/L^{1.25-1.75}$.}
\label{fig:efield}
\end{figure}
\begin{figure}
\epsfxsize=3.25in\epsfbox{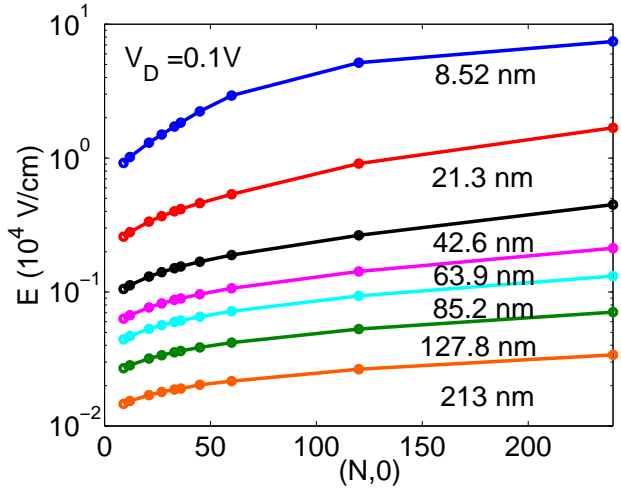}
\caption{The electric field at the mid point versus 
nanotube chirality index for various lengths at an applied bias of 100 mV.
 The electric field increases approximately as 
$N^{0.25-0.75}$.}
\label{fig:efieldN}
\end{figure}
While the total density of states  (DOS) per unit length of
metallic nanotubes is independent of diameter, the DOS per subband  
and free charge carried by the crossing subbands are
inversely proportional to the diameter.  
As a result, we find that the
screening in metallic nanotubes degrades with diameter. 
The potential
drop for  nanotubes with diameters of 0.94 nm [(12,0) nanotube],
2.82 nm [(36,0)] and 4.7 nm [(60,0)] are shown in Fig. \ref{fig:pot2}.
Clearly, screening is poorer in the larger diameter nanotube. In fact,
while the potential drops by 45 mV in a distance of 1 nm from the edge
for the (12,0) nanotube, the potential drops for (36,0) and (60,0) 
 are only 37 mV and 31 mV respectively. 
The inset of Fig. \ref{fig:pot2} shows a 
substantially larger electric field away from the edges of the large
diameter nanotube.  
Another difference between large and small diameter 
nanotubes is due to the different geometry contained in the 
electrostatic Green's function (\ref{eq:green}).
Higher number of atoms per ring in 
larger diameter tubes slightly offsets  
the deterioration of screening due lower density of states.

Fig. \ref{fig:dens2} shows the self-consistent  charge profiles, corresponding to the 
potential in Fig. \ref{fig:pot2}.  The interplay between the competing
factors of different density of states 
and electrostatic Green's function results
in different charge profiles. In the larger diameter tubes more charge is accumulated at the edges.

The electric field at the center of the nanotube as a function of 
length is shown in Fig. \ref{fig:efield} for the tube with a diameter of
0.94 nm. We find that for all diameters, the electric field decreases
more rapidly than $\frac{1}{L}$, where $L$ is the length of the
nanotube. The exact power law however depends on the diameter.
If the computed electric field versus length is fit to $\frac{1}{L^a}$,
the exponent $a$ increases with increase in diameter. The value of
$a$ increases from 1.25 to 1.75 as the diameter increases from 0.94
to 18.85 nm.
The dependence of the electric field on the chirality index
 of the tube is shown in Fig. \ref{fig:efieldN}.
The electric field versus diameter  ($D=aN/\pi$) 
can be fit to $D^{b}$, where $b$ is in the range of 0.25 and 0.75. 
The presence of non zero electric field inside a conducting nanotube is not 
totally surprising: it is a consequence of a lower electron density of states
 in 1D.

\subsection{Transport}

\begin{figure}
\epsfxsize=3.25in\epsfbox{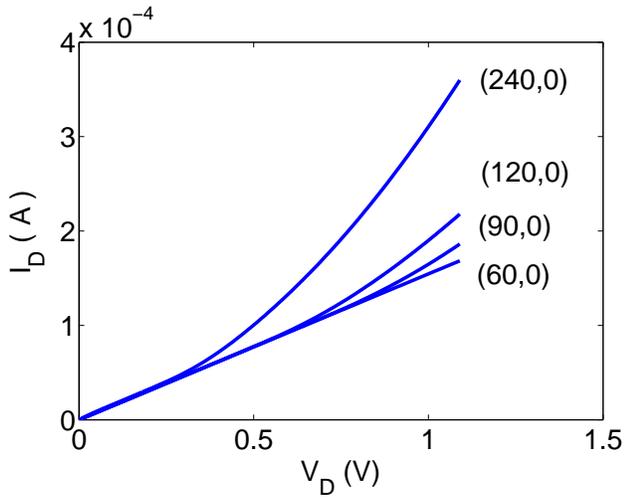}
\caption{The current-voltage characteristics of large diameter nanotubes in the ballistic limit.
The current shows faster than linear increase with bias due to the contribution of non crossing subbands.
}
\label{fig:IVbal}
\end{figure}

\begin{figure}
\epsfxsize=3.25in\epsfbox{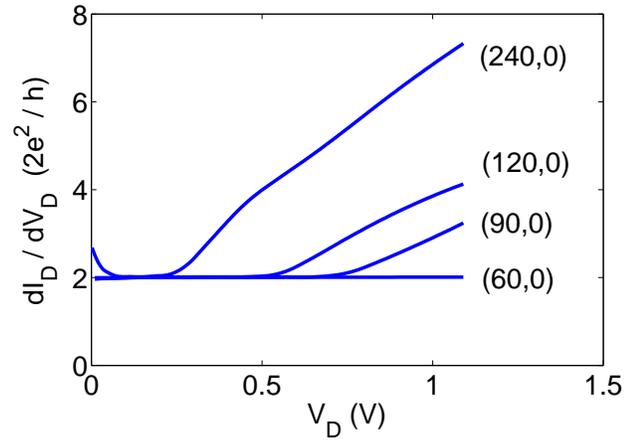}
\caption{The differential conductance as a function 
of bias  in the ballistic case. 
Contribution of non crossing subbands in (240,0) nanotube leads to 
the non monotonic behavior of conductance. High bias increase 
corresponds to Zener tunneling and is determined by the bandgap of the lowest non crossing subband.}
\label{fig:dIdV}
\end{figure}

We now discuss the current-voltage characteristics as a function of
bias for small and large diameter nanotubes of length 42.6 nm, in the
ballistic limit. The ballistic current and differential conductance 
for the different diameters are  shown in Fig.\ref{fig:IVbal} and
Fig.\ref{fig:dIdV}.
For the moderate diameter nanotubes ($N<90$), 
only two crossing subbands contribute to transport.
The  current increases linearly
with applied bias and the differential conductance is $4e^2/h$.

\begin{figure}
\epsfxsize=3.25in\epsfbox{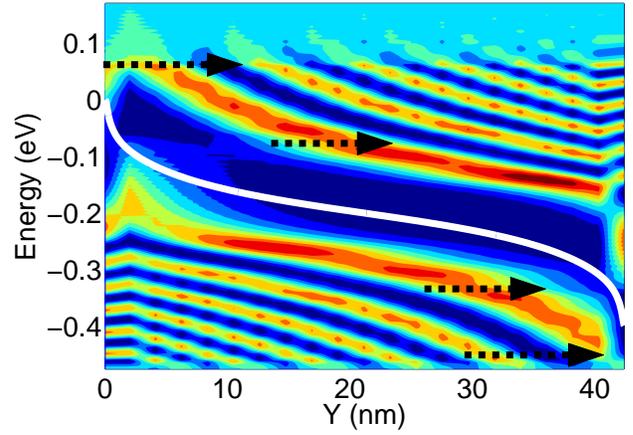}
\caption{Local density of states in the lowest non crossing 
subband of a (240,0) nanotube
under a bias of 0.4 V as a function of position and energy.
Brighter color shows higher DOS.
 Also shown is the potential profile.
The arrows correspond to 
four contributions to a current due to the non crossing 
subband in Fig.\ref{fig:CURvsE}}.

\label{fig:DOSvsE}
\end{figure}

In large diameter nanotubes ($N =$ 90, 120 and 240),
ballistic current shows super linear increase.
The conductance of large diameter nanotubes  starts 
increasing after a threshold bias.
This increase occurs due to Zener tunneling contribution of 
the non crossing subbands: when the bias 
becomes larger than twice the bandgap of the lowest non crossing subband,
electrons can tunnel from  valence band 
states ($E<-E^{non}_{g}/2+V(y)$) in the source to  
the conduction band  states ($E >E^{non}_{g}/2+V(y)$) in the middle of the tube (the channel) and also
from  valence band 
states in the channel  to the conduction band  states in the drain.
The lowest non crossing subbands ($q=N/3 \pm 1$) in (90,0), (120,0) and (240,0) 
have bandgaps $E^{non}_{g}=$ 331, 249 and 125 meV respectively 
and start to contribute 
to current at biases of twice these values.
In the ballistic limit, the self-consistently
calculated current of (240,0) nanotube  at 1 V is 310 $\mu$A and the differential conductance 
is almost $13 e^2 / h$, signifying that a large number of subbands are
contributing to current. As the barrier width at energies corresponding to
 valence band edge in the source $E_v^S$ and conduction band edge in the drain $E_c^D$ 
decreases with bias, the transmission and Zener tunneling current increase.
The density of states (DOS) in the lowest non crossing subband in (240,0) 
nanotube at a bias of 0.4 V is shown in Fig. \ref{fig:DOSvsE}. Note the 
increased DOS in the bandgap at energies $E_v^S$ and $E_c^D$. 

Another qualitative difference occurs at zero bias in (240,0) nanotube (Fig.\ref{fig:dIdV}).
The zero bias conductance of (240,0) nanotube
is larger than $4e^2/h$ because the 
non crossing subbands  are partially 
filled and contribute to current:
the first non crossing subband opens at an energy of 2.4 $kT$ 
from the band center.
This contribution is a simple intraband transport, 
determined by the population of the conduction band in the source and the 
valence band in the drain, but rather insensitive to the details of the potential distribution.
At slightly higher biases
the non crossing subbands  contribution  to current 
saturates to a constant value and the contribution to differential conductance decreases to zero, while
the total conductance decreases to $4e^2/h$ . 
\begin{figure}
\epsfxsize=3.25in\epsfbox{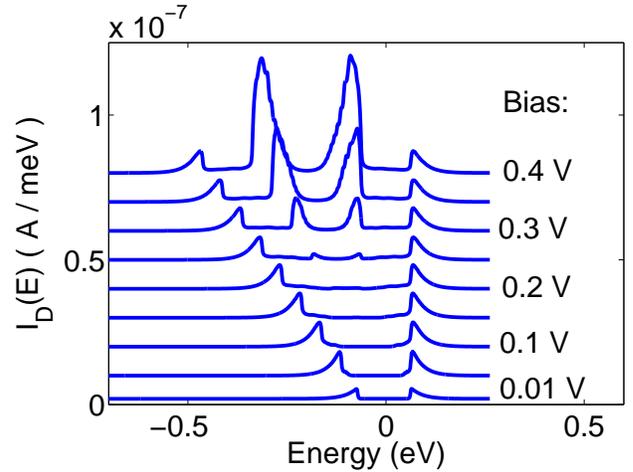}
\caption{A set of current density 
versus energy curves due to the lowest non crossing subband in a (240,0) nanotube under different biases.
Two outermost peaks correspond to transport in the valence and conduction bands.
Two inner peaks correspond to Zener tunneling between valence and conduction band states.
}
\label{fig:CURvsE}
\end{figure} 
 
To elucidate the effect of non crossing subbands
further, Fig.\ref{fig:CURvsE} shows
the evolution with bias of 
current versus electron energy in the lowest  non crossing subband.   
At low biases there are two peaks in the current density corresponding to
partially filled valence and conduction bands. 
The height of the peaks increases only slightly and 
then saturates at all higher biases. These peaks in
 the current density are responsible for the increased value of zero bias conductance.
When  bias is higher than twice the bandgap, 
two new peaks in the current density occur at $E_v^S$ and $E_c^D$,
corresponding to  Zener tunneling at the source and drain respectively.
The height of these peaks increases rapidly due to the exponential dependence on the barrier width.

The important consequence of the different origin of these 
four peaks is the temperature dependence of differential conductance. 
At low temperatures, the outermost peaks will vanish,
reducing the value of conductance to $4e^2/h$. However, the height and shape of the inner peaks are 
 insensitive to the temperature which implies the robustness of Zener tunneling.

In conclusion, we find that 
in the ballistic (low bias/large diameter) regime, applied 
bias drops mostly at the edges. 
However, the 1-dimensionality of the nanotube leads to very  
slow ($\sim 1/L^{\alpha}$) screening of the electric 
field as compared to a bulk crystal.
In agreement with 
experimental measurements of Ref. \cite{bachtold-prl-00}, 
our calculations show larger electric field
for larger diameter nanotubes. 
The conductance of large diameter nanotubes shows 
non monotonic bias dependence: at low and high bias the 
conductance is higher than $4e^2/h$ due to the 
contribution of non crossing subbands. 

We are grateful to Dr. Hatem Mehrez for his
help with  
the complex contour integration of DOS.

\end{document}